\documentclass[two column,rsi,amsmath,amssymb,superscriptaddress] {revtex4-1}
\usepackage{color,graphics}
\usepackage{graphicx}
\usepackage[sort&compress]{natbib}
\usepackage{hhline}
\usepackage{soul}
\usepackage{xcolor}
\usepackage{bm}

\usepackage[normalem]{ulem}
\begin{document}
\setcounter{totalnumber}{3}

\newcommand{\BTA}{\mathrm{(BTA)}_2\mathrm{PbBr}_4}
\newcommand{\PEA}{\mathrm{(PEA)}_2\mathrm{PbBr}_4}

\title{Vibrational relaxation dynamics in layered perovskite quantum wells}

\author{Li Na Quan}
 \thanks{These two authors contributed equally}
 \affiliation{Department of Chemistry, University of California, Berkeley}
 \affiliation{Materials Science Division, Lawrence Berkeley National Laboratory}

\author{Yoonjae Park}
 \thanks{These two authors contributed equally}
 \affiliation{Department of Chemistry, University of California, Berkeley}

\author{Peijun Guo}
 \affiliation{Center for Nanoscale Materials, Argonne National Laboratory}

\author{Mengyu Gao}
 \affiliation{Materials Science Division, Lawrence Berkeley National Laboratory}
 \affiliation{Department of Materials Science and Engineering, University of California, Berkeley}

\author{Jianbo Jin}
 \affiliation{Department of Chemistry, University of California, Berkeley}
 
\author{Jianmei Huang}
 \affiliation{Department of Chemistry, University of California, Berkeley}

\author{Jason K. Copper}
 \affiliation{Chemical Science Division, Lawrence Berkeley National Laboratory}

\author{Adam Schwartzberg}
 \affiliation{Molecular Foundry, Lawrence Berkeley National Laboratory}
 
\author{Richard Schaller}
 \affiliation{Center for Nanoscale Materials, Argonne National Laboratory}
     
\author{David T. Limmer}
 \email{dlimmer@berkeley.edu}
 \affiliation{Department of Chemistry, University of California, Berkeley}
\affiliation{Materials Science Division, Lawrence Berkeley National Laboratory}
\affiliation{Chemical Science Division, Lawrence Berkeley National Laboratory}
\affiliation{Kavli Energy NanoScience Institute, Berkeley, California, Berkeley}

\author{Peidong Yang}
 \email{p_yang@berkeley.edu}
 \affiliation{Department of Chemistry, University of California, Berkeley}
\affiliation{Materials Science Division, Lawrence Berkeley National Laboratory}
\affiliation{Chemical Science Division, Lawrence Berkeley National Laboratory}
\affiliation{Kavli Energy NanoScience Institute, Berkeley, California, Berkeley}
\affiliation{Department of Materials Science and Engineering, University of California, Berkeley}

\date{\today}

\begin{abstract}
Organic-inorganic layered perovskites are two-dimensional quantum wells with layers of lead-halide octahedra stacked between organic ligand barriers. The combination of their dielectric confinement and ionic sublattice results in excitonic excitations with substantial binding energies that are strongly coupled to the surrounding soft, polar lattice. However, the ligand environment in layered perovskites can significantly alter their optical properties due to the complex dynamic disorder of soft perovskite lattice. Here, we observe the dynamic disorder through phonon dephasing lifetimes initiated by ultrafast photoexcitation employing high-resolution resonant impulsive stimulated Raman spectroscopy of a variety of ligand substitutions. We demonstrate that vibrational relaxation in layered perovskite formed from flexible alkyl-amines as organic barriers is fast and relatively independent of the lattice temperature. Relaxation in aromatic amine based layered perovskite is slower, though still fast relative to pure inorganic lead bromide lattices, with a rate that is temperature dependent. Using molecular dynamics simulations, we explain the fast rates of relaxation by quantifying the large anharmonic coupling of the optical modes with the ligand layers and rationalize the temperature independence due to their amorphous packing. This work provides a molecular and time-domain depiction of the relaxation of nascent optical excitations and opens opportunities to understand how they couple to the complex layered perovskite lattice, elucidating design principles for optoelectronic devices.
\end{abstract}

\maketitle

Organic-inorganic hybrid layered perovskite quantum wells have optoelectronic properties that can be adapted to enable a diverse set of applications including solar cells, light-emitting diodes, semiconductor lasers, and photodetectors\cite{mitzi1994conducting, saparov2016organic, nnano2014149, science1228604, nmat4271, natrevmats2017}. The structural stability and degree of quantum confinement can be tuned by varying the inorganic semiconductor, organic barrier compositions and stoichiometry independently, making them more viable for many device applications than their bulk counterparts\cite{chondroudis1999,chen20182d,yuan2016perovskite,stoumpos2016ruddlesden,dhanabalan2020directional}. The photoluminescence quantum efficiency of layered perovskites in particular can be altered by varying the molecular configuration of the organic ligands, and as a result, various compositions have been used for both light absorbing and light emitting applications\cite{gong2018electron,dou2015atomically,gao2019molecular}. The local distortions of the inorganic octahedra in layered perovskites create a complex energetic landscape for charges that activate additional scattering mechanisms, and determine emergent optoelectronic properties such as exciton and carrier transport, and light emission\cite{acsnanolett8b04276}. However, the precise nature of electron lattice interactions in these materials remains to be understood. Femtosecond lasers have made it possible to impulsively generate and detect coherent phonons in semiconductor nanostructures with time-resolved pump and probe measurements. Time resolved photocarrier dynamics in quantum wells and semiconductor superlattices can provide detailed insight into the relevant interaction mechanisms between coherent phonons and coherently prepared electronic wave packets.

Here we employ ultrafast pump-probe transient absorption spectroscopy on 2D layered perovskites to investigate the direct dynamic interplay of optically generated excitons in the perovskite layer with their surrounding organic lattice. Our observations show a phonon dephasing process with a strong dependence on the organic barrier. In conjunction with molecular dynamics simulations, we find that the composition of organic ligands and inorganic quantum well thickness can substantially change the dephasing rate of optical phonons and their temperature dependence, due to varying degrees of anharmonicity in the lattice and dynamic structural disorder. This molecular and time domain insight into optical relaxation sheds light on the emergent electron-lattice interactions in these materials and enables their design for optoelectronic devices.

\begin {figure}
\centering\includegraphics [width=8.5cm] {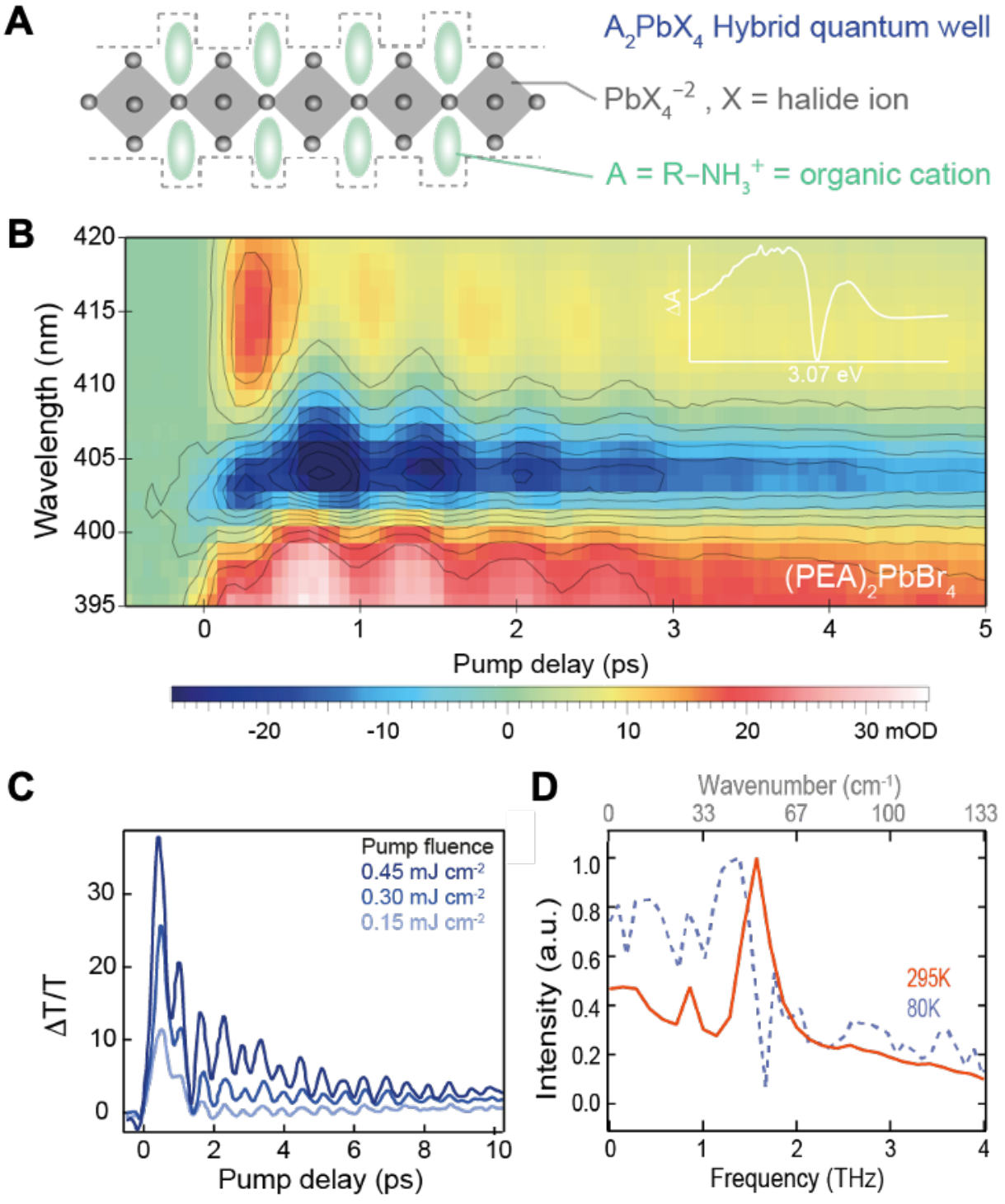}
\caption{
Impulsive generation of coherent phonon oscillations in $\PEA$.
(A) Schematic illustration of the hybrid quantum-well structure. 
(B) Time-resolved differential transmission (dT) spectrum from $\PEA$.
(C) Extracted coherent phonon oscillations from time-resolved differential transmission spectrum measured with different pump intensity at 80K. Probe energy of 3.0 eV.
(D) Fourier transform spectrum of the coherent oscillation measured at room temperature (red solid lines) and 80 K (blue dashed lines).
}
\label{Fig1}
\end{figure}

We studied thin films and single crystals of A$_2$PbX$_4$ (A$=$R$-$NH$_3^+$) that can be synthesized by mixing precursors at desired stoichiometric ratios, followed by spontaneous self-assembly of the quantum well structure, illustrated in Fig.~\ref{Fig1}A (see details in Methods). This class of materials is known to exhibit moderate quantum confinement effects, resulting in narrow-band emission (bandwidth eV) combined with exceptionally large oscillator strength ($7\times 10^{-2} \textrm{cm}^{-1}$)\cite{mauck2019excitons}. Optical phonons are expected to be the most strongly coupled modes to nascent electronic excitations in these materials as such optical phonons will modulate charge transport and Coulomb screening\cite{thouin2019phonon, c6mh00275g, acsjpclett6b01425}. Such motions can in principle be investigated by Raman scattering by examining the lineshapes which informs the lifetime, enabling the elucidation of the roles of confinement, extended interfaces, and disorder on the modes\cite{lowph}. However, these are often difficult to disentangle in highly complex systems which have inherently coupled broadening mechanisms.

By employing a narrow optical excitation and following the excitation in the time domain from the associated transient modulation of the reflectivity, we can decouple the effects from homogeneous and inhomogeneous broadening in the Raman scattering lineshape. Femtosecond optical excitation allows for an impulsive coherent electronic excitation that generates coherent phonons as it relaxes, as it is shorter than the inverse of the phonon period\cite{hase2003birth}. Since the generation and subsequent relaxation processes are strongly affected by the coupling of phonon modes to the photoexcited states, the real-time observation of coherent phonons can offer crucial insight into to the dynamic electron-phonon coupling.

To detect the coherent phonons, we use a standard pump–probe configuration in which the observed differential transmission (dT) is modulated due to changes in the complex dielectric function from electron-phonon coupling.  Thus, the dephasing time can be characterized by the decay of the dT modulation amplitude. Using time-resolved measurements, it has been possible to monitor the dephasing of coherent optical phonons near $q=0$ in polar semiconductors such as GaP\cite{PRBbron}, GaAs\cite{PRBvallee1}, ZnSe\cite{PRBbron}, InP\cite{PRBvallee2}. In Figs. \ref{Fig1}B and C, we show that the dT dynamics in the thin films following resonant excitation with a sub-50-fs visible pulse (3.26 eV). The vibrational coherence map can be resolved due to the spectrally dispersed probe-wavelength and collected as a function of pump-probe delay time and probe wavelength. We observed pronounced oscillations resulting from optical phonons on top of an exponentially decaying background.  The background signal is related to rapid electronic excited state relaxation. In bulk perovskites such as CsPbX$_3$, a coherent phonon induced oscillation is less visible due to the decrease in both phonon mode amplitude and dephasing rate at room temperature\cite{PRBnemec}. The dT oscillations can be assigned to two different phonon modes with frequency calculated by fast Fourier transform of 0.85 THz (wavenumber) and 1.57 THz (wavenumber), as shown in Fig. \ref{Fig1}D. We also measured the photo-excited phonon dynamics at various pump wavelengths with both off-resonance and resonant excitations (3.5- 2.9 eV). We observed oscillatory response in all cases \textcolor{black}{(Fig. S6)}, indicative of a strong coupling of optical excitation to the lattice.

\begin {figure}
\centering\includegraphics [width=8.5cm] {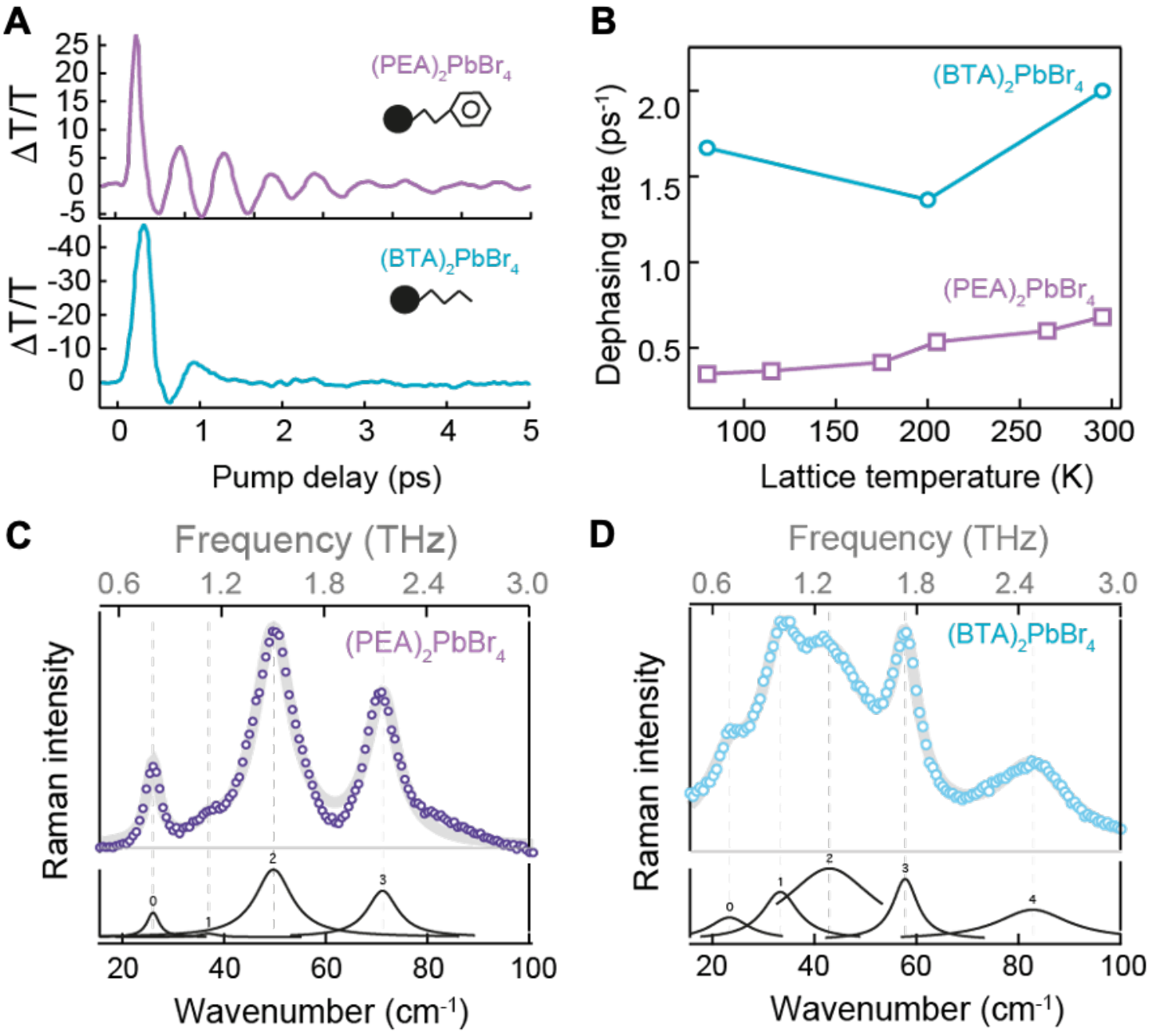}
\caption{
Structural information of hybrid perovskite quantum wells with different organic ammonium cation as barriers and the corresponding phonon-lattice dynamics. (A) Resonant impulsive Raman spectra by pump-probe measurement with quantum wells packed with different geometry at room temperature. TA is plotted at the probe energy of 3.0 eV. (B) Lattice temperature dependent dephasing rate of coherent optical phonon dynamics. Non-Resonant Raman spectra (fitted with Lorentzian) of (C) $\PEA$ perovskite quantum well and (D) $\BTA$ perovskite quantum well.
}
\label{Fig2}
\end{figure}

We considered one ligand derived from an alkyl group, n-butylamine (BTA), and one from an aromatic group, phenylethylamine (PEA). The packing geometry of the organic barriers leads to a structural deformation of inorganic octahedral \textcolor{black}{(Fig. S2)}, and thus strongly affects the energy, lifetime, and localization of the band-edge exciton, and as a result, affects changes to the electrical transport properties of these materials\textcolor{black}{\cite{JPCLstraus,Naturehu}}. We observed a clear trend in photoluminescence efficiency, where the perovskites with aromatic organic groups typically shown high photoluminescent yield of up to 50 \% at room temperature while those made of alkyl groups show much lower yields. Therefore, a comprehensive understanding of their dynamical structure under photoexcitation is important for the development of an efficient materials platform. The stable phase of $\PEA$ is a lattice with a triclinic space group (P1), and the PEA organic barriers stack in a T-shape arrangement via strong $\pi-\pi$ interaction, forming relatively rigid crystal geometry. The $\BTA$ perovskites, on the other hand, form a lattice with a Pbca space group, with the alkyl-group organic cations generating a weak quantum well-to-well stacking interaction indicated by the relatively broad diffraction peaks \textcolor{black}{(Fig. S2)}. Fig. \ref{Fig2}A reports the ultrafast resonant impulsive Raman probed at 3.0 eV for both $\PEA$ and $\BTA$, where the oscillation spectra have been obtained after subtraction of an exponential decay of carriers. This oscillation observed is reproducible for different positions on the sample. Upon photoexcitation, we observe qualitatively different dynamics depending on the organic barriers. For $\PEA$, we find persistent oscillations over 5 ps, while for $\BTA$, oscillations are dephased rapidly and any mode assignment becomes uncertain. The corresponding ultrafast phonon-dephasing rate extracted was significantly diminished in the perovskites with the alkyl group, relative to the aromatic group. These measurements were repeated at different temperatures, spanning 80 K to 300 K. Additionally, we also varied alkyl chain lengths, and the shape of aromatic groups and measured their corresponding phonon dephasing dynamics \textcolor{black}{(Fig. S7)}.

The spectral fringe visibility of the resulting optical phonon vibrations, as a function of time delay, is used to measure the phonon dephasing time. This is extracted and plotted as a function of temperature in Fig. \ref{Fig2}B. In $\PEA$, it showed a strong temperature-dependent phonon dephasing rate, indicative of the anharmonicity of optical phonon coupled to the electronic states. The coherent phonon population in $\PEA$ perovskites decays within 4-5 ps at room temperature, while the dephasing time can be extended up to 10 ps at 80 K. This increase of the phonon relaxation rate with temperature is anticipated by the increase phonon-scattering enabled by anharmonic effects in the vibrational potential energy\cite{PRBbalkanski,PRBverma}. In $\BTA$, with alkyl chain group, the dephasing rate was around 4 times faster than the aromatic ligand. Further the dephasing rate in the alkyl ligand had no significant temperature dependence. In bulk crystalline semiconductors, a temperature independent phonon linewidth is usually explained as consequence of an elastic scattering of coherent phonons by a perturbing potential of lattice defects\cite{APLhase}. However, as we discuss below, in this case it is a high degree of dynamic rather than static disorder that results in a suppressed temperature dependence in $\BTA$. Given a common inorganic framework, the different temperature dependence observed for the different cations indicates that the organic ligands play an important role in determining the lifetime of the modes most strongly coupled to the photogenerated excitons.

\begin {figure*}
\centering\includegraphics [width=15.5cm] {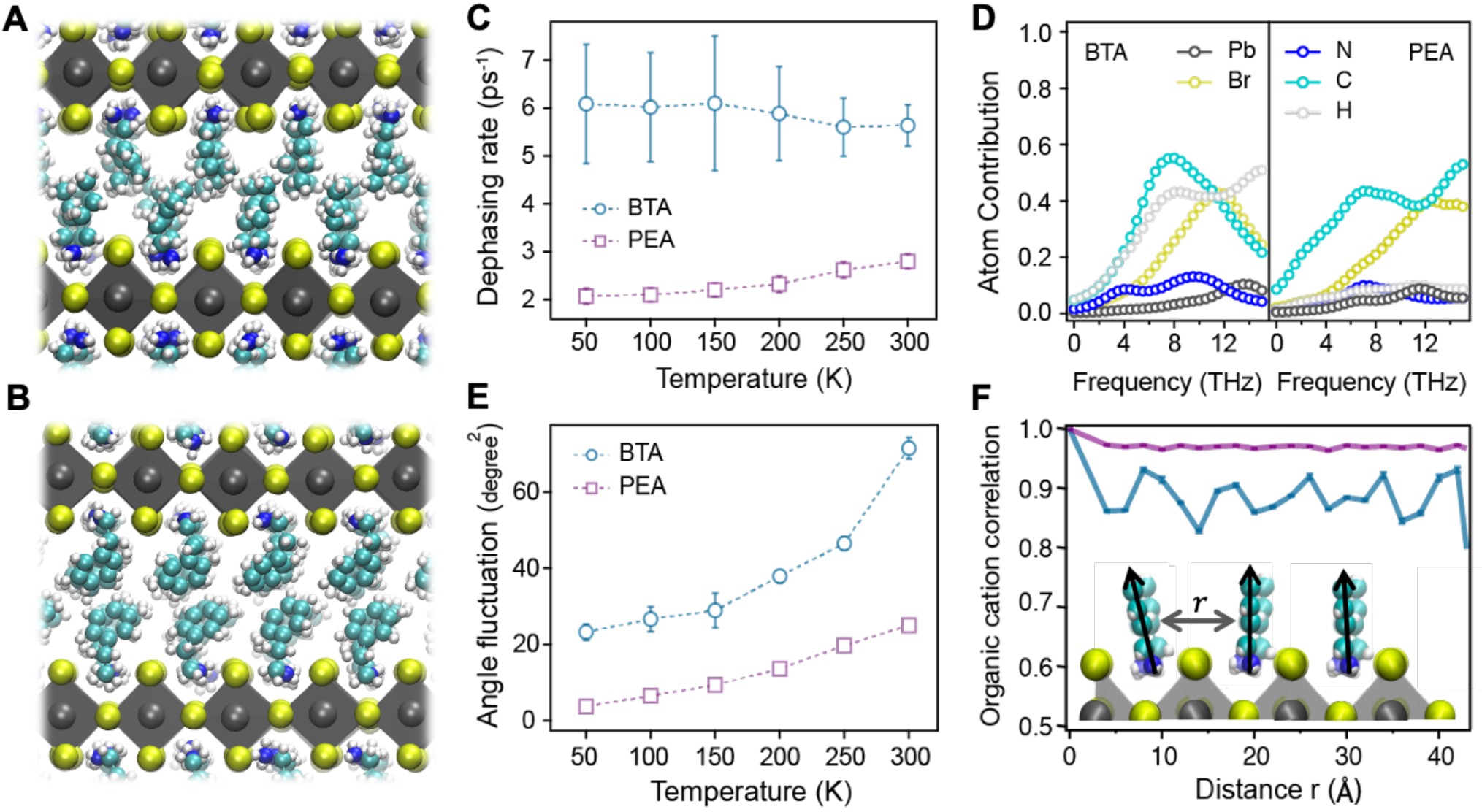}
\caption{Simulation of vibrational dynamics. (A,B) Snapshots of simulations on $\BTA$ (A) and $\PEA$ (B) perovskites. Colored dots below are the color schemes for the snapshots. 
(C) Dephasing rate of the lowest optical mode in $\BTA$ (blue circles) and $\PEA$ (purple squares).  
(D) Contributions from each atom type to the vibrational mode as a function of frequencies of vibrational mode in each layered perovskite. 
(E) Fluctuations of angles between vertical axis and NC vector (schematically defined in Fig. S14) at each temperature. 
(F) Normalizaled spatial correlations between the nitrogen-carbon dipole vector (black arrows) as a function of the lateral distance between vectors. Dotted lines in (C) and (E) are guides to the eye.}
\label{Fig3}
\end{figure*}

To gain initial insight into the optical phonons and their structural disorder, we further performed non-resonant Raman scattering measurements with all perovskites we mentioned previously. The low-frequency (10$-$100 $\textrm{cm}^{-1}$) Raman scattering is expected to be associated with vibrational information of inorganic octahedral, and thus uniquely suited to measure the distortion induced modulated structures. We observed in Fig. \ref{Fig2}C, three strong symmetric vibrational scattering feature below 100 cm$^{-1}$ in $\PEA$, which previously assigned to various vibrational modes of PbBr$_6$ octahedral frameworks. Especially the strongest vibrational scattering at 55 cm$^{-1}$ is well matched with optical phonon oscillation frequency that we observed from time-resolved optical measurements. Surprisingly, we found a significant broadening and spectral shift of inorganic phonon modes in perovskites with alkyl group $\BTA$, shown in Fig. \ref{Fig2}D, which is characteristically associated with inhomogeneous distortion in the octahedral lattice. The Raman signal at 40 cm$^{-1}$ is a sum of two broadened symmetric Lorentzian subpeaks of Pb-Br coordination, which means the Pb-Br bond lengths in $\BTA$ are distorted significantly compare with the $\PEA$. The Raman result implying that the variable orientation of the different organic cations is largely responsible for the lattice distortion.

To obtain an atomistic perspective on the vibrational dynamics of $\BTA$ and $\PEA$, we performed molecular dynamics simulations. To simulate each perovskite, we employed an empirical model \textcolor{black}{(Tables S1-3)} with fixed point charges whose reduced computational cost enables the study of extended system sizes over the long times required to average over the slow fluctuations of the ligands\cite{ffref,mattoni}. While simple, the model reasonably reproduces experimental values of the lattice constants and mechanical properties of these and related materials \textcolor{black}{(Table S4)}. Snapshots of the simulations with the different organic cations are shown in Fig. \ref{Fig3}A and \ref{Fig3}B. To study the dynamics of vibrational relaxation, we extract the phonon modes by computing the lattice Green’s function and solving the associated eigenvalue equation from the dynamical matrix. We employ the fluctuation dissipation theorem to compute the dynamical matrix from the displacement correlations from a simulation at 50 K, and extract the effective vibrational frequencies and phonon modes \textcolor{black}{(Eq. S6)} at that temperature\cite{LKong1,Dove}. It is expected that optical phonons with lower energy are attributed to the bending and rocking motions of PbX$_3$ inorganic framework\cite{lowph,dahod2020low,nagai2018longitudinal}. While we confirm that they participate in these modes, in our calculations we found that the organic cations contribute significantly to the low- frequency modes in the layered perovskites. The participation ratio for each atom type as a function of frequency is shown in Fig. \ref{Fig3}D, where for both PEA and BTA cations the organic ligand atoms contribute up to 40\% to the longitudinal and transverse optical modes.

For the lowest frequency optical mode in each layered perovskite, we computed the dephasing rate over a range of temperatures from 50 K to 300 K. 
In the classical limit, the Fermi's golden rule rate for vibrational relaxation, $\Gamma_\lambda$, can be computed by the Fourier transform of the force-force correlation function of the optical mode\cite{egorov1999quantum},
$$
\Gamma_\lambda = \frac{1}{k_\mathrm{B}T} \int_0^\infty dt \cos(\omega_\lambda t) \langle F_\lambda(0)F_\lambda(t) \rangle
$$
where $F_\lambda$ is the force on the $\lambda$'th mode due to the anharmonic coupling to the surrounding lattice and $\omega_\lambda$ its corresponding frequency. The average $\langle .. \rangle$ is taken over an isobaric, isothermal ensemble with 1 atm of pressure and varying temperature. 
The classical treatment of the nuclei is justified by the fact that the energy for the optical phonon is much less than the thermal energy, $\hbar \omega_\lambda \ll k_\mathrm{B}T$, where $\hbar$, $k_\mathrm{B}$, and $T$ are Planck's constant, Boltzmann's constant and the temperature respectively~\cite{skinner2001calculating}. 
 In agreement with experiment, Fig. \ref{Fig3}C  shows the dephasing rates in $\BTA$  are higher than the rates for $\PEA$  across the range of temperatures studied. Further consistent with the experiment, we find a strong temperature dependence of the dephasing rate with the PEA cation, while the dephasing rate with the BTA cation is insensitive to the temperature. 
%

\begin {figure}
\centering\includegraphics [width=8.4cm] {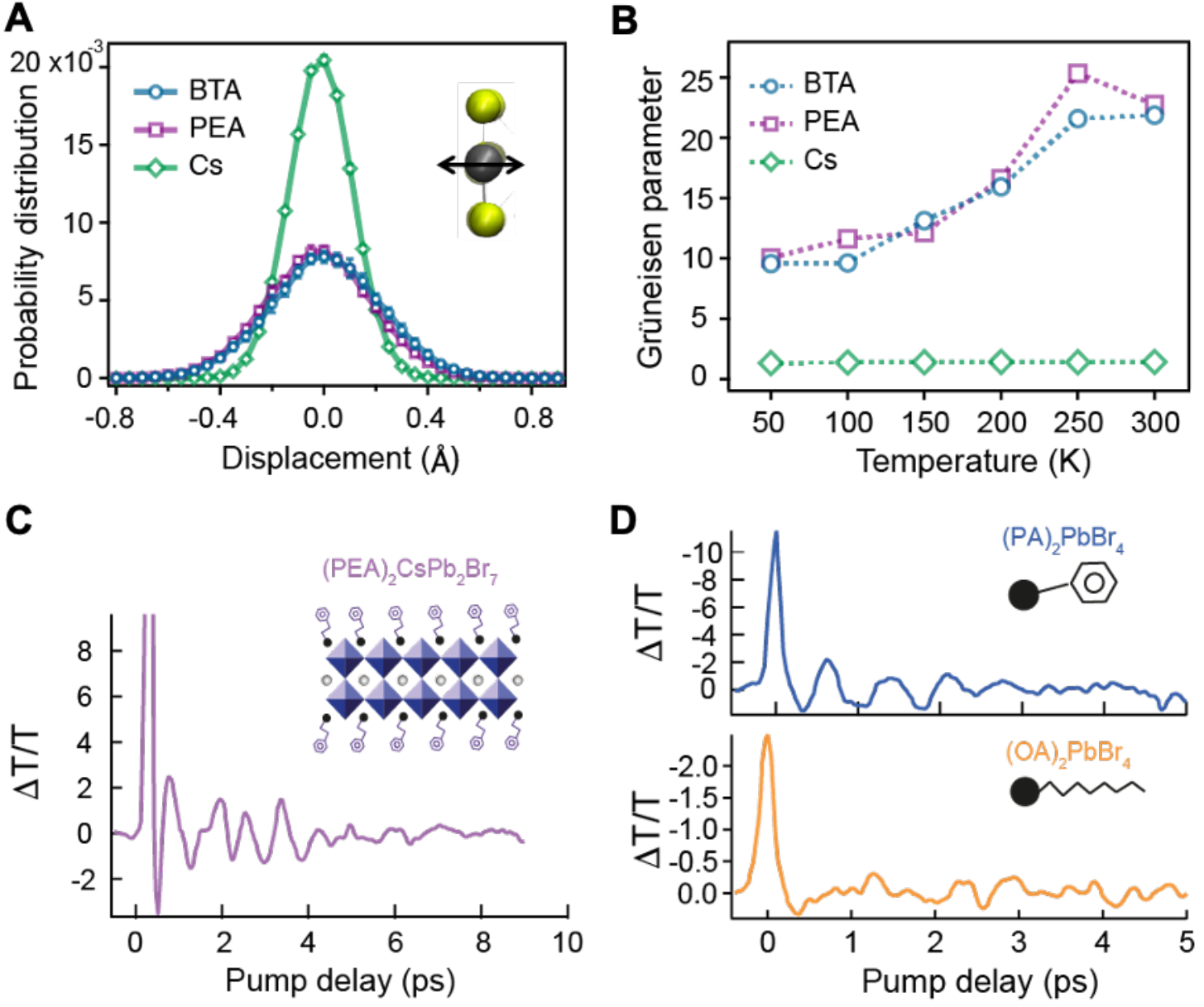}
\caption{Generality and mitigation of coherent phonon dephasing dynamics. (A) Probability distribution of Pb atom displacement from its average position in (BTA)$_2$PbBr$_4$ (blue circles), (PEA)$_2$PbBr$_4$ (purple squares), and CsPbBr$_3$ (green diamonds) perovskites. 
(B) Grüneisen anharmonicity parameters in layered and bulk perovskites. Dotted lines are used to connect neighboring symbols.
Oscillatory response from coherent phonon dynamics of \textcolor{black}{(C)} $n=2$ multi-quantum well perovskites with PEA
organic cations and (D) $n=1$ quantum well perovskites with phenylammonium (\textcolor{black}{top}) and octylammonium (\textcolor{black}{bottom}) organic cations.}
\label{Fig4}
\end{figure}

We find that the high phonon dephasing rates in both $\PEA$ and $\BTA$  are a consequence of significant anharmonic coupling of the optical mode to the ligand barriers that increase phonon scattering. We have characterized the anharmonicity both directly through structural measures and indirectly through the dependence of the modes on temperature and volume.
To quantify anharmonicity structurally within the perovskites, we analyzed the angle distribution of each organic cation.
Fig. \ref{Fig3}E shows the mean squared fluctuations of the angle, $\langle \delta \Theta^2\rangle$, between the axis perpendicular to the inorganic octahedra plane, $\hat{\mathbf{z}}$, and the nitrogen-carbon unit vector, $\mathbf{R}_\mathrm{NC}$, as a function of each temperature, such that $\cos(\Theta)=\mathbf{R}_\mathrm{NC} \cdot \hat{\mathbf{z}}$. If the local potential of the ligand orientation were harmonic, according to the fluctuation-dissipation theorem the fluctuations would increase linearly as the temperature is increased. For both the PEA and BTA cations, we find a nonlinear temperature dependence, with a more significant departure from linearity in BTA relative to PEA. The corresponding distribution, \textcolor{black}{(Fig. S14)}, is significantly non-Gaussian for BTA, while only marginally so for PEA at 300 K. Both have optical frequencies that show a strong temperature dependence \textcolor{black}{(Fig. S11)}, while for BTA the optical mode becomes significantly mixed with increasing temperature \textcolor{black}{(Fig. S10)}. 
Given the large contribution of the ligand to the optical mode in both materials, the stronger anharmonicity of $\BTA$ than in $\PEA$ explains the higher rate of phonon relaxation in BTA perovskite as resulting from stronger phonon scattering.

The relative temperature independence of $\BTA$ can be understood by noting the significant dynamic disorder in that lattice relative to  $\PEA$. This is quantified by the spatial correlation between nitrogen-carbon dipoles, $\bm{\mu}$,  
$$
C(r) = \langle \bm{\mu}(0) \cdot \bm{\mu}(r)\rangle
$$
separated by a lateral distance $r$. This correlation function is illustrated in Fig. \ref{Fig3}F, normalized by its value at the origin, $C(r)/C(0)$. 
For $\PEA$, the normalized correlation function, which ranges from 0 to 1, is unstructured, and does not decay appreciably over 40 $\mathrm{\AA}$. For $\BTA$, however, the correlation decreases within a unit cell, and drops periodically to as low as 0.8. Compared to PEA, the lower long-ranged correlation in BTA details an amorphous packing of BTA even at low temperature. 
Rather than the expected phonon-defect scattering, this shows that the persistent dynamic disorder results in a temperature insensitive relaxation rate. Molecularly this results from the stronger, directional $\pi-\pi$ stacking of PEA \textcolor{black}{as shown in Fig.~\ref{Fig3}B}, relative to the weaker isotropic van der Waals interactions of BTA. 

We have compared these simulation findings to a CsPbBr$_3$ bulk perovskite, by considering two additional measures of anharmonicity. At the molecular level, we analyze the structure of the inorganic framework. We have computed the displacement distribution of the lead atoms from their equilibrium lattice positions at 300 K. 
Fig. \ref{Fig4}A illustrates that the distribution is much narrower in CsPbBr$_3$ perovskite relative to $\PEA$ or $\BTA$, indicating that structures of layered perovskites are soft and less structurally rigid as compared to bulk perovskite, while the distributions are nearly identical for both ligands. Additionally, we have calculated the Grüneisen  parameter, \textcolor{black}{$\gamma_\lambda = -\partial \ln \omega_\lambda/ \partial \ln V$}, for the $\lambda$'th mode with lattice volume $V$ (\textcolor{black}{Fig. S13}). Since the frequency of harmonic modes does not depend on the volume and temperature, a higher value of Grüneisen parameter indicates a higher strength of anharmonicity. 
\textcolor{black}{Fig.~\ref{Fig4}B} shows $\gamma$ for the lowest optical model as a function of temperature. We find that the anharmonicity in layered perovskites is much higher than in the bulk perovskite, consistent with our expectation, and the former grows with temperature. Using an Umklapp scattering model~\cite{ziman2001electrons}, the Grüneisen parameter and the corresponding lower shear moduli of the layered perovskites are qualitatively able to reproduce the ordering of the dephasing rates of  $\BTA$ and $\PEA$, and suggest that both are an order of magnitude larger than that expected for CsPbBr$_3$ \textcolor{black}{(Fig. S12)}.

While the direct comparison to the bulk CsPbBr$_3$ perovskite is not possible experimentally due to the decreased phonon amplitude and facile charge separation, we have studied the vibrational dynamics of lattices with increased inorganic layers. We synthesized a single crystal (PEA)$_2$CsPb$_2$Br$_7$ ($n = 2$) perovskite and measured its ultrafast coherent optical phonon dynamics. The transient absorption signal is shown in  \textcolor{black}{Fig. \ref{Fig4}C} in which the dephasing lifetime is longer, 8 ps, compared with single layered $\PEA$ ($n = 1$) counterpart. This is consistent with mitigating the anharmonic coupling to the ligand barriers, and subsequent lower phonon-to-phonon scattering rates. Additionally, as a direct test of the underlying molecular packing argument emerging from the simulations, we synthesized and measured the coherent phonon dynamics of (PA)$_2$PbBr$_4$ where PA is Phenylammonium and (OA)$_2$PbBr$_4$ where OA is Octylammonium perovskites. Fig. 4D demonstrates that the phonon lifetime is longer with the PA organic cation, as clarified by its persistent oscillatory response, as compared to the OA organic cation, in which no oscillations are observed. This is consistent with the ability of aromatic cations to leverage $\pi-\pi$ interactions and stabilize a more ordered ligand layer relative to akyl cations, supporting the previous analysis of the BTA and PEA organic cations. 

We have presented an experimental and theoretical description of vibrational relaxation in layered perovskites. We studied the dephasing dynamics of two different organic cations in depth and observed a difference in lifetime and response to the temperature of the phonons, implying that organic cations play an important role in determining the relaxation dynamics coupled to photogenerated excitations. This work elucidates that anharmonicity and dynamic disorder from organic cations facilitates vibrational relaxation and optical vibrational modes are largely mixed with the organic species. Our work contributes to the deeper understanding of phonon dynamics and may provide the foundation for future studies on the mechanism of electron-phonon interaction in Ruddlesden-Popper phase perovskites.


\begin{acknowledgments}
This work was supported by the U.S. Department of Energy, Office of Science, Office of Basic Energy Sciences, Materials Sciences and Engineering Division, under Contract No. DE-AC02-05-CH11231 within the Physical Chemistry of Inorganic Nanostructures Program (KC3103). This research used resources of the National Energy Research Scientific Computing Center, a U.S. Department of Energy Office of Science User Facility operated under Contract No. DE-AC02-05CH11231. Simon Teat and Laura McCormick are acknowledged for helping to accommodate the single-crystal X-ray beamtime. Use of the Center for Nanoscale Materials, an Office of Science user facility, was supported by the U.S. Department of Energy, Office of Science, Office of Basic Energy Sciences, under Contract No. DE-AC02-06CH11357. The Raman system acquisition was supported by the NSF MRI proposal (EAR-1531583).  
\end{acknowledgments}

\bibliography{phonon_draft_final}
\end{document}